\documentclass[%
 reprint,
 superscriptaddress,
 amsmath,amssymb,
 aps,
 pra,
]{revtex4-2}

\usepackage{graphicx}
\usepackage{dcolumn}
\usepackage{bm}
\usepackage{color}
\usepackage{hyperref}
\hypersetup{colorlinks, linkcolor=green, urlcolor=blue, citecolor=green}

\begin{document}

\title{Isotopic variations and Zeeman-like splitting in the spectra of nonlinear photonic meta atoms}

\author{S. Zhang}
\author{I. Babushkin}
\author{U. Morgner}
\author{A. Demircan}
\author{O. Melchert}
\email{melchert@iqo.uni-hannover.de}

\affiliation{Leibniz Universit\"at Hannover, Institute of Quantum Optics,  Welfengarten 1, 30167 Hannover, Germany}

\affiliation{Leibniz Universit\"at Hannover, Cluster of Excellence PhoenixD, Welfengarten 1A, 30167 Hannover, Germany}

\date{\today}

\begin{abstract}
We study photonic meta-atoms, a unique class of composite solitary wave, 
supported in nonlinear waveguides.
%
%
We establish an analogy to one-dimensional soft-core atoms, allowing to describe the complex dynamics via concepts from atomic physics.
%
%
%
%
Higher-order dispersive effects cause specific spectral resonances characteristic for the eigenspectrum of a meta-atom.
We  demonstrate that subtle changes in this level spectrum causes frequency shifts of the resonances.
These shifts consist of isotopic and isomeric contributions that can be distinguished in terms of a simple model.
We further demonstrate a generic mechanism that causes a Zeeman-like splitting of resonance lines.
\end{abstract}

\maketitle

\paragraph*{Introduction.} 
%

%
Solitons describe self-confined solutions of nonlinear wave-equations, traveling with constant shape and speed~\cite{Zabusky:PRL:1965}. 
%
%
Their stability is enabled by a balance of linear and nonlinear effects, and
they occur in various fields of physics, such as, e.g., 
hydrodynamics \cite{Russell:BAR:1844}, 
plasma dynamics \cite{Zabusky:PRL:1965}, 
and nonlinear optics \cite{Mitschke:BOOK:2016}.
The study of solitons and their interactions thus is an interdisciplinary endeavour, addressing theoretical concepts from all these fields.
%
%
%
%
%
Their complex dynamics has further motivated formal analogies to quantum physics in terms of the Ramsauer-Townsend effect~\cite{Belyaeva:EPJD:2012}, Geiger-Nuttal law~\cite{Serkin:JMO:2013}, and nuclear reactions~\cite{Serkin:OC:2023}.
%
%
In nonlinear optics, a generic soliton model is the nonlinear Schrödinger equation (NSE) \cite{Mitschke:BOOK:2016}.
Perturbation of the NSE facilitates further intriguing effects such as optical Cherenkov radiation \cite{Akhmediev:PRA:1995}.
Specifically, higher orders of dispersion can support alternate anomalous-normal-anomalous dispersion ranges, enabling soliton spectral tunneling \cite{Serkin:EL:1993}. 
%
%
%
Corresponding dispersion characteristics have been realized in nonlinear waveguides~\cite{Khallouf:OE:2025}, hollow-core fibers~\cite{Zeisberger:SR:2017}, and photonic crystal fibers~\cite{Willms:PTL:2023}.
Such higher-order NSE (HONSE) models can exhibit, e.g., 
generalized dispersion Kerr solitons \cite{Tam:PRA:2020,Lourdesamy:NP:2021}, 
two-frequency soliton molecules  \cite{Melchert:PRL:2019,Melchert:OL:2021}, and, photonic meta-atoms \cite{Melchert:OPTIK:2023,Melchert:OL:2023,Gutierrez:OL:2025}.
%
%
%
Among these, meta-atoms consist of a soliton and weak trapped pulses, bound together via a cross-phase modulation induced trapping potential \cite{Melchert:PRL:2019}.
They are defined by a Schrödinger-type eigenvalue problem, formulated on a temporal domain, ``hidden'' within the HONSE.
%

%
Here, we demonstrate that earlier meta-atom analogies to one-dimensional (1D) quantum dynamics \cite{Melchert:PRL:2019,Melchert:OPTIK:2023,Gutierrez:OL:2025} can be substantiated with a view to 1D soft-core atoms in strong-field physics \cite{Javanainen:PRA:1988,Eberly:JOSAB:1989}.
%
%
They are established as theoretical tools to study, e.g., atomic ionization \cite{Eberly:JOSAB:1989,Su:PRL:1990}, and electron capture \cite{Grobe:PRA:1991}. 
%
%
%
In the considered HONSE, higher-order dispersive effects couple a meta-atoms trapped states to the continuum.
This defines a decay mechanism that does not exist in quantum mechanics.
%
Yet, we show that the analogy to strong-field physics can be used to explain the complex propagation dynamics by comparing it to familiar concepts from atomic physics. 
Specifically, we demonstrate that slight changes in the trapped-state structure cause frequency shifts on the resonance lines, allowing to draw an analogy to isotopic and isomeric shifts, i.e.\ spectroscopic tools to determine volume and charge distribution effects on atomic spectra \cite{King:BOOK:1984,Berengut:N:2025}.
We further demonstrate a generic radiation mechanism, known from oscillating solitary waves \cite{Driben:OE:2013,Melchert:SR:2020,Melchert:NJP:2023}, that causes a Zeeman-like splitting of the resonance lines.

\begin{figure}[t!]
\centering{\includegraphics[width=\linewidth]{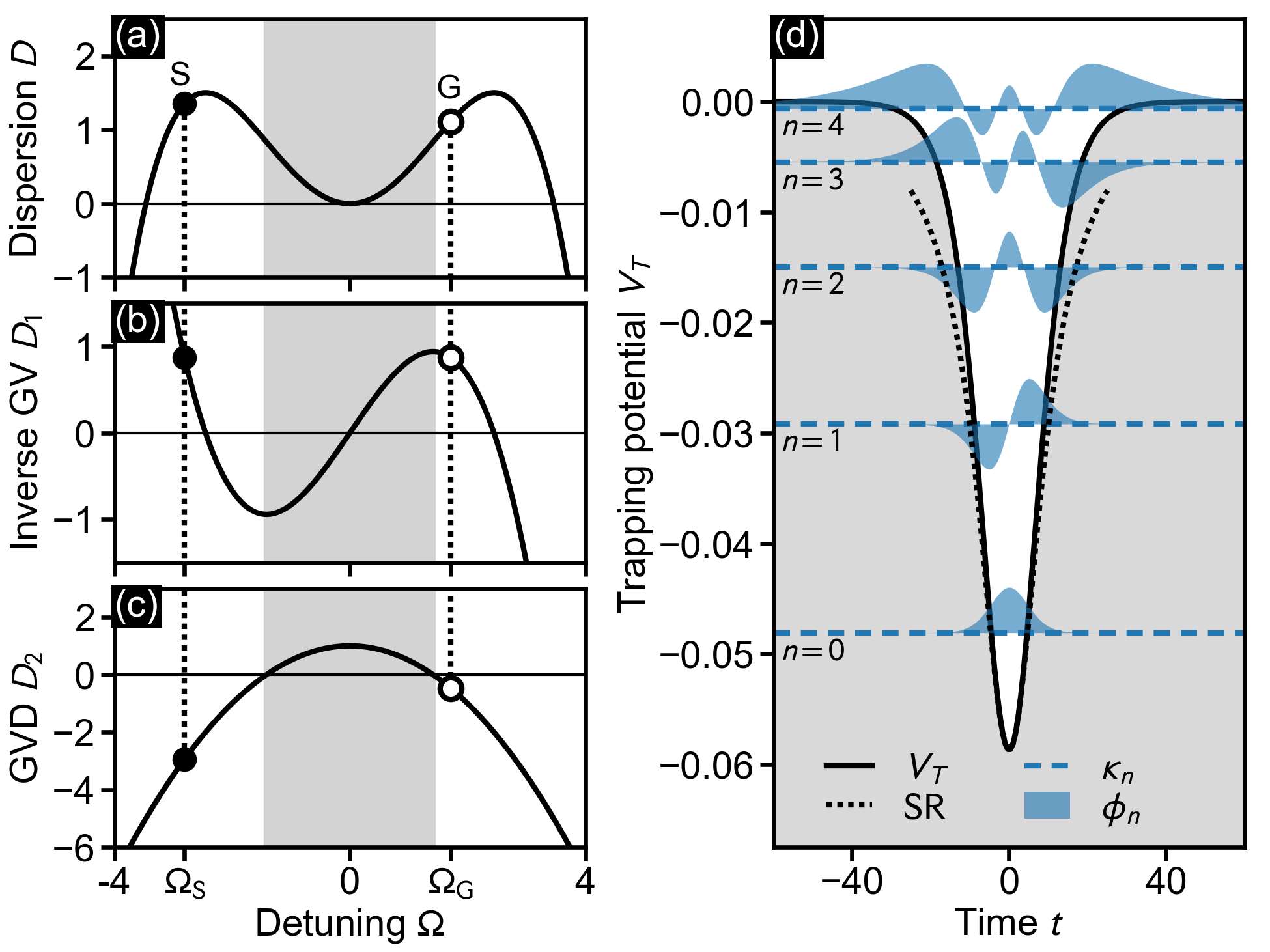}}
\caption{
Details of the model.
(a) Dispersion,
(b) inverse group-velocity, and
(c) group-velocity dispersion.
Domain of normal dispersion [$D_2(\Omega)>0$] is shaded gray.
Dot and circle refer to the loci $\Omega_{\rm{S}}$ and $\Omega_{\rm{G}}$ of the soliton (S) and Gaussian pulse (G), respectively.
%
%
(d) Trapping potential $V_{\rm{T}}$ with trapped states $\phi_n$ at eigenvalues $\kappa_n$ for $n=0,\ldots,4$. 
Dotted line (SR) denotes the short-range approximation of $V_{\rm{T}}$ (see text). 
\label{fig:01}}
\end{figure}

\paragraph*{Model and Methods.}
%
%
%
We consider a complex-valued envelope $A\equiv A(z,\tau)$, governed by the HONSE
\begin{align}
\label{eq:HONSE}
 i\partial_z A =
 \left(
   \frac{\beta_2}{2}\,\partial_\tau^2
 - \frac{\beta_4}{24}\,\partial_\tau^4
 \right) A - \gamma |A|^2 A , 
\end{align}
with nonlinear coefficient $\gamma$ (units: ${\rm{W^{-1}\,m^{-1}}}$), and quadratic and quartic dispersion, $\beta_2>0$  (units: ${\rm{ps^2\,m^{-1}}}$) and $\beta_4<0$  (units: ${\rm{ps^4\,m^{-1}}}$), respectively.
%
%
For brevity we neglect physical units and set parameters to $\beta_2=1$, $\beta_4=-1$, and $\gamma=1$.
Using the spectral derivative identity  $\left[(i\partial_\tau)^n -\Omega^n\right]\, e^{-i\Omega \tau}=0$ for the detuning $\Omega$ (units: ${\rm{rad}\,ps^{-1}}$), gives the frequency-domain representation $D(\Omega)=\frac{\beta_2}{2}\Omega^2+\frac{\beta_4}{24}\Omega^4$ of the dispersion, see Fig.~\ref{fig:01}(a).
Figures~\ref{fig:01}(b,c) show the corresponding  inverse group-velocity (GV) $D_1\equiv\partial_\Omega D$, and group-velocity dispersion (GVD) $D_2\equiv\partial_\Omega^2 D$.
The dispersion features two separate domains of anomalous dispersion [$D_2<0$], with zero-dispersion points $\Omega_{\rm{Z1},\rm{Z2}}=\mp\sqrt{2}$, and supports GV matching across a vast frequency gap [Fig.~\ref{fig:01}(b)].
%
%
%
This HONSE admits photonic meta-atoms \cite{Melchert:PRL:2019,Melchert:OL:2023,Gutierrez:OL:2025,Melchert:OPTIK:2023}, 
i.e.\ composite solitary waves of the form
\begin{align}
\label{eq:ansatz}
 A(z,\tau)=A_{\rm{S}}(z,\tau)e^{-i(\Omega_{\rm{S}} \tau+\beta_{0,{\rm{S}}}z)}  +A_{\rm{G}}(z,\tau)e^{-i(\Omega_{\rm{G}} \tau + \beta_{0,{\rm{G}}}z)},
\end{align}
for two GV matched, narrowband pulses $A_{\rm{S}}$ and $A_{\rm{G}}$ with frequency loci $\Omega_{\rm{S}}$ and $\Omega_{\rm{G}}$ in separate domains of anomalous dispersion, and $\beta_{0,{\rm{S/G}}}=D(\Omega_{\rm{S/G}})$.
%
%
Assuming $\max(|A_{\rm{G}}|)\ll \max(|A_{\rm{S}}|)$ allows to reduce Eq.~(\ref{eq:HONSE}) to the coupled system  \cite{Melchert:PRL:2019,Melchert:OPTIK:2023}
\begin{subequations}\label{eq:coupled_set}
\begin{align}
i\partial_z \,A_{\rm{S}} &= -\frac{|\beta_{2,\rm{S}}|}{2}\partial_t^2 A_{\rm{S}}
- \gamma |A_{\rm{S}}|^2 A_{\rm{S}},\label{eq:NSE}\\
i\partial_z \,A_{\rm{G}} &= -\frac{|\beta_{2,\rm{G}}|}{2}\partial_t^2 A_{\rm{G}} - 2\gamma |A_{\rm{S}}|^2 A_{\rm{G}}.\label{eq:SE}
\end{align}
\end{subequations}
Therein $t=\tau - z/v_0$ is a retarded time that accounts for the GV $v_0$ of both pulses, and $\beta_{2,{\rm{S/G}}}=D_2(\Omega_{\rm{S/G}})$. 
System~(\ref{eq:coupled_set}) neglects higher orders of dispersion, and is linearized about the weak pulse $A_G$.
%
%
Equation~(\ref{eq:NSE}) represents a standard NSE with fundamental soliton $A_{\rm{S}}(z,t)=\sqrt{P_0}\,{\rm{sech}}(t/t_0)\,e^{i\kappa_{\rm{S}}z}$ with $P_0=|\beta_{2,{\rm{S}}}|/(\gamma\,t_0^2)$ and $\kappa_{\rm{S}}=\gamma P_0/2$.
%
Equation~(\ref{eq:SE}), in which pulse coupling is achieved by cross-phase modulation, specifies a Schrödinger equation for a weak pulse $A_{\rm{G}}$, describing a fictitious particle of mass $m=|\beta_{2,{\rm{G}}}|^{-1}$ subject to a soliton-induced trapping potential $V_{\rm{T}}(t)=-2\gamma P_0 \,{\rm{sech}}^2(t/t_0)$.
%
%
In terms of the characteristic wavenumber \mbox{$K_0= |\beta_{2,{\rm{G}}}|\,t_0^{-2}$} and the auxiliary parameter 
$\nu=-\frac{1}{2} + (\frac{1}{4} + 4 \left|\beta_{2,\rm{S}}/\beta_{2,\rm{T}} \right|)^{1/2}$
it can be written as
\begin{align}
V_{\rm{T}}(t) = -\frac{K_0}{2}\frac{\nu\,(\nu+1)}{{\mathrm{cosh}}^2\left(t/t_0\right)} \approx -\frac{K_0}{2}\frac{\nu\,(\nu+1)}{1+\left(t/t_0\right)^2},\label{eq:VS}
\end{align}
%
where the last expression states its short-range approximation for $t\ll t_0$ [Fig.~\ref{fig:01}(d)].
%
%
%
%
%
%
The ansatz $A_{\rm{G}}(z,t)=\phi(t)\,e^{-i\kappa z}$ turns Eq.~(\ref{eq:SE}) into the meta-atomic Schrödinger eigenproblem \mbox{$\kappa \phi = -(2m)^{-1}\partial_t^2\phi + V_{\rm{T}}(t) \phi$}.
%
%
It can be solved exactly~\cite{Lekner:AJP:2007,Melchert:PRL:2019}, giving \mbox{$N=\lfloor \nu \rfloor + 1$} stationary trapped states $\phi_n$ at wavenumber eigenvalues \mbox{$\kappa_n=-\frac{1}{2}K_0\,(\nu-n)^2$}, labeled by the integer index $n = 0,\ldots, \lfloor \nu \rfloor$~\cite{Melchert:PRL:2019,Melchert:OPTIK:2023}.
%
%
%
%
%
These trapped states comprise direct optical analogues of quantum mechanical bound states \cite{Melchert:PRL:2019}. 
%
%
Composite pulses that satisfy Eqs.~(\ref{eq:ansatz}-\ref{eq:coupled_set}) define photonic meta-atoms.
%
%
%
An example for $t_0=10$, $\Omega_{\rm{S}} = -2.81$, and $\Omega_{\rm{G}} = 1.71$ [Figs.~\ref{fig:01}(a-c)], defining a meta-atom with $\nu=4.525$ and $N=5$, is shown in Fig.~\ref{fig:01}(e).
To refer to a meta-atom with parameters $\nu$ and $t_0$, we below use the concise notation ``$(\nu=4.525,t_0=10)$-nuclide''.

%

%
%
Given the inherent 1D nature of meta-atoms, which are manifested in the temporal domain $t$, it is intriguing to report a formal analogy to 1D soft-core atoms in strong-field physics \cite{Javanainen:PRA:1988,Eberly:JOSAB:1989}, which are manifested in a 1D spatial domain $x$. 
%
%
In atomic units, the underlying soft-core Coulomb potential with charge $Z$ and softening parameter $a$ reads
%
\begin{align}
V_{\rm{SC}}(x) = - \frac{Z}{\sqrt{x^2 + a^2}} \approx -\frac{Z}{a} \frac{1}{1+(x/\sqrt{2}a)^2}, \label{eq:VSC}
\end{align}
where the last expression states the short-range approximation for \mbox{$x\ll a$}.
%
%
%
%
%
%
%
%
%
%
In quantum dynamic simulations, $Z$ and $a$ are commonly adjusted to match the GS energy of a selected atom \cite{Sallai:PRA:2024}.
%
%
Here, we proceed by instead matching the well depths in Eqs.~(\ref{eq:VS},\ref{eq:VSC}), allowing to formulate an analogy between meta-atoms and soft-core atoms.
%
%
Introducing meta-atomic units, i.e.\ setting $K_0=1$ and $t_0=1$, the well depth in Eq.~(\ref{eq:VS}) is $V_{\rm{T}}(0)=-\frac{1}{2}\nu(\nu+1)$.
%
%
Selecting $Z$ and $a=2/Z$ in Eq.~(\ref{eq:VSC}) allows to match depths for the effective charge $\nu^\prime=-\frac{1}{2} + \tfrac{1}{2}\sqrt{1 + 4 Z^2} \approx Z -\tfrac{1}{2}$ for all $Z\geq1$.
%
%
%
Below we associate the discrete number of bound states $N$ with the atomic number $Z$ of a soft-core atom.
Moreover, we associate the soliton duration $t_0$ with the extent of an atomic nucleus, related to its mass number.
We point out that meta-atoms with $N-1\leq \nu <N$ exhibit the same number of trapped states.
Adopting nuclear physics terminology \cite{Gold:BOOK:1997}, we refer to meta-atoms with equal $N$ but different $t_0$ as ``isotopes''.
%
%
%
%

%

%
%
Subsequently, we perform pulse propagation simulations of Eq.~(\ref{eq:HONSE}) in terms of fixed-stepsize and adaptive-stepsize schemes \cite{Melchert:CPC:2022}, see Appendix~\ref{appendix:b}.
%
%
%
For the supporting theoretical analysis it is useful to state a variant of the meta-atomic Schrödinger Eq.~(\ref{eq:SE}) in which higher orders of dispersion and self-phase modulation are retained, given by 
\begin{align}
i\partial_z A_{\rm{G}} &= \left[-D_{\rm{T}}(i\partial_t)  + V_{\rm{T}}(t) -\gamma |A_{\rm{G}}|^2 \right] A_{\rm{G}}, \label{eq:GPE}
\end{align}
with 
$D_{\rm{T}}(i\partial_t) = \frac{\beta_{2,\rm{G}}}{2}(i\partial_t)^2 + \frac{\beta_{3,\rm{G}}}{6}(i\partial_t)^3 +\frac{\beta_{4,{\rm{G}}}}{24}(i\partial_t)^4$, and $\beta_{n,{\rm{G}}}=\partial_\Omega^nD|_{\Omega=\Omega_{\rm{G}}}$.
%
%
This Gross-Pitaevskii-type equation allows to predict resonant frequencies, excited due to higher orders of dispersion, via a phase-matching analysis \cite{Yulin:OL:2004}.
Specifically, a trapped state with eigenvalue $\kappa_n$ drives resonant radiation at a frequency $\Omega_{{\rm{R}},n}$, defined by the wavenumber matching condition $-\kappa_n=D_{\rm{T}}(\Omega_{{\rm{R}},n})$, see Appendix~\ref{appendix:a}.
%
%
%
The resonance loci for each $(\nu,t_0)$-nuclide are specific, comprising a kind of ``spectral fingerprint''. 


\begin{figure}[t!]
\centering{\includegraphics[width=\linewidth]{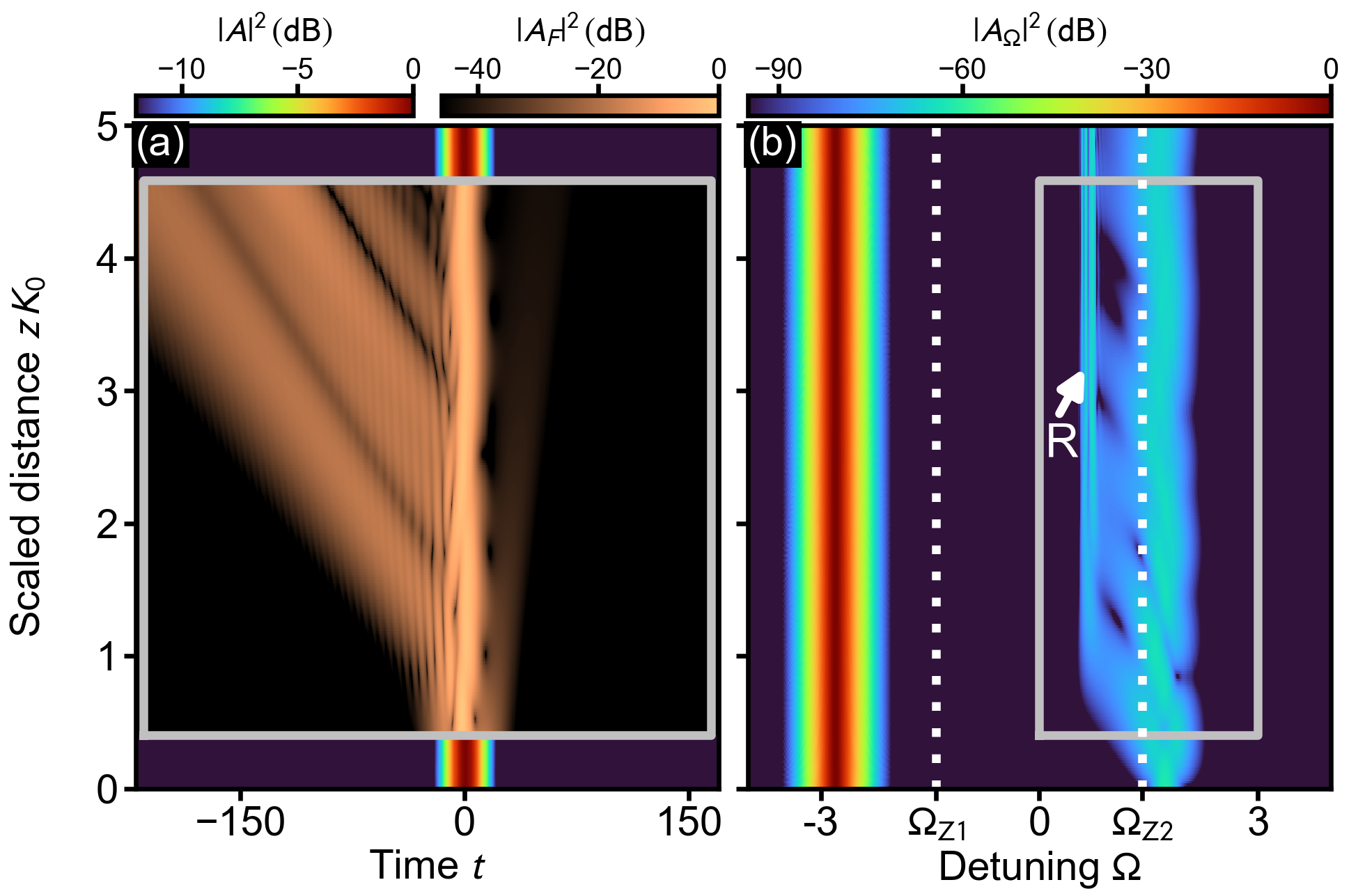}}
\caption{
Propagation dynamics demonstrating the excitation of resonant radiation.
(a) Time-domain propagation dynamics of a soliton, and a superimposed Gaussian pulse (see text).
%
%
(b) Corresponding spectrum.
Dotted lines indicate zero-dispersion points.
The arrow (labeled R) indicates the location of resonantly excited modes.
%
%
%
The field $A_F$, enclosed by the box in (a), is defined by the spectrum enclosed by the box in (b).
\label{fig:02}}
\end{figure}

\begin{figure}[t!]
\centering{\includegraphics[width=\linewidth]{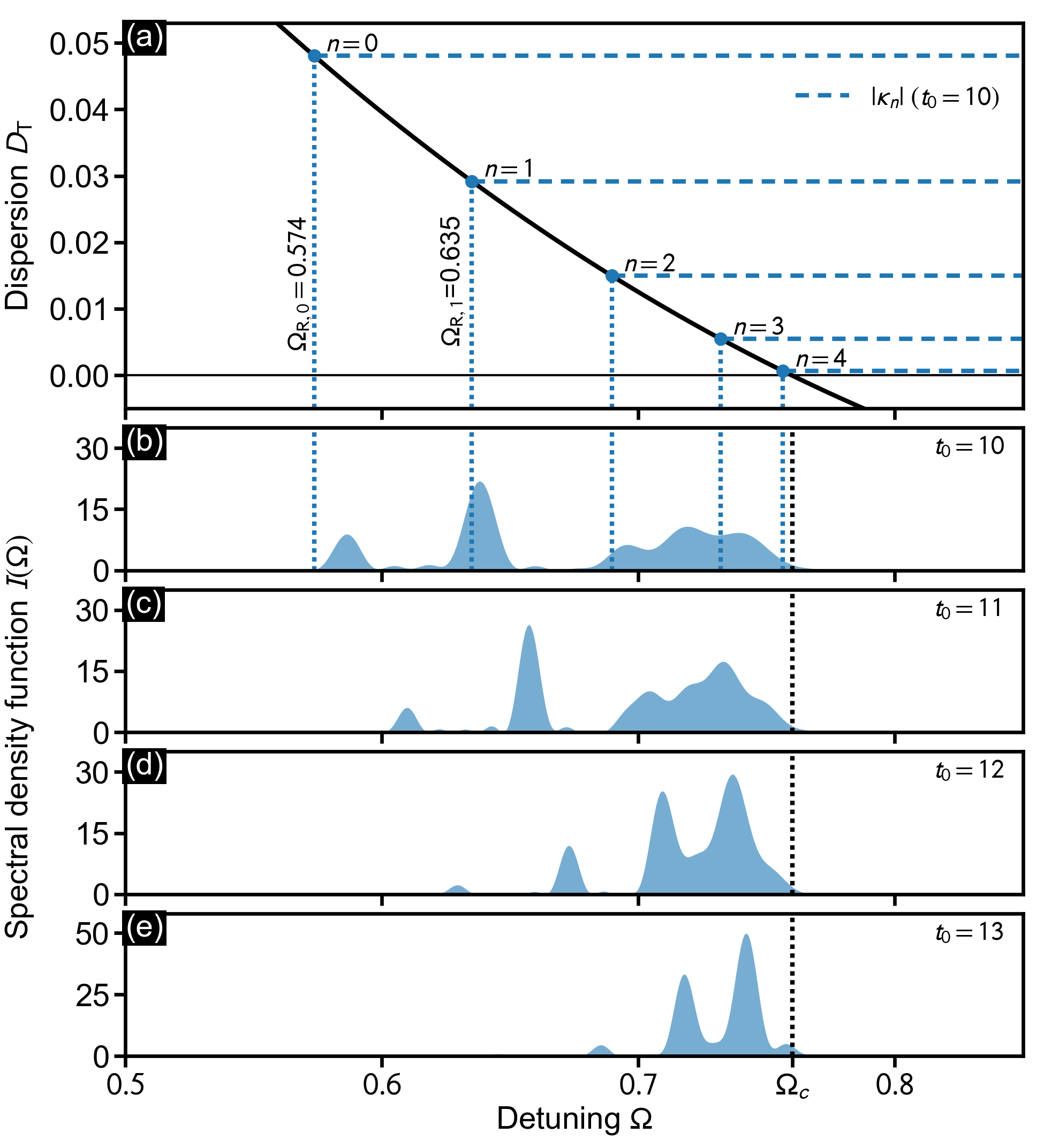}}
\caption{Spectral signature of meta-atom isotopes.
(a) Graphical solution of the phase-matching condition (see text).
%
%
(b-e) Resonance spectral lines for selected values of $t_0$.
(b) $t_0=10$, (c) $t_0=11$, (d) $t_0=12$, and, (e) $t_0=13$.
%
%
The spectral density function (SDF) is normalized as $\int_{0}^{\Omega_c}I(\Omega)~{\rm{d}}\Omega=1$.
Black dotted line in (b-e) indicates cutoff frequency $\Omega_c$.
Blue dotted lines in (b) indicate predicted resonances at $\Omega_{{\rm{R}},n}$ for $n=0,\ldots,4$.
\label{fig:03}}
\end{figure}

\paragraph*{Results.}
%
%
%
%
The association to strong-field physics and atomic physics explain the complex propagation dynamics, goverened by Eq.~(\ref{eq:HONSE}), by comparing it to familiar concepts. 
In this context, we demonstrate an effect analog to the isotopic shift \cite{King:BOOK:1984,Berengut:N:2025}, referring to spectroscopic differences found among nuclides with equal atomic number but different mass number.
%
%
%
%
In Fig.~(\ref{fig:02}) we show the propagation dynamics of a fundamental soliton ($\Omega_{\rm{S}}=-2.81$, $t_0=10$) and a superimposed pulse in the form $A_0(t)=\sqrt{P_0}\,{\rm{sech}}(t/t_0)\,e^{-i\Omega_{\rm{S}}t} + A^\prime(t)\,e^{-i\Omega^\prime t}$, governed by Eq.~(\ref{eq:HONSE}).
The soliton induces the trapping potential of the $(4.525,10)$-nuclide, shown in Fig.~\ref{fig:01}(d).
Its stationarity, presumed by Eq.~(\ref{eq:NSE}), can be verified in Figs.~\ref{fig:02}(a,b).
While the choice of $A^\prime$ is not crucial, it should obey $\max(|A^\prime|)\ll \sqrt{P_0}$ and not be orthogonal to any of the trapped states. 
We therefore opt for a Gaussian pulse $A^\prime(t)=10^{-3}\sqrt{P_0} \exp(-t^2/2t_{{\rm{G}}}^2)$ with $t_{{\rm{G}}}=10$ at $\Omega^\prime=1.72$, having a slight GV mismatch relative to the soliton.
This allows for a non-vanishing overlap with even and odd trapped states, exciting all states upon propagation.
%
%
%
As soon as energy is captured by the trapped states, phase-matched resonant modes are excited, indicated by the arrow in Fig.~\ref{fig:02}(b).
%
%
%
In the time-domain, the resonant radiation and the meta-atom exhibit a large GV mismatch [Fig.~\ref{fig:02}(a)].
This emission of resonant radiation results in a decay of the energy trapped by the meta-atom~\cite{Babushkin:Preprint:2022}.
%
%
%
%
%
The single-sided emission in Fig.~\ref{fig:02}(a) is caused by the dispersion profile $D_{\rm{T}}$.
Its shape is sensitive to the frequency location $\Omega_{\rm{G}}$, and it controls the transfer of energy from the trapped states to dispersive waves.
%
%
A graphical solution of the underlying resonance condition is shown in Fig.~\ref{fig:03}(a),
%
%
%
predicting the trapped groundstate ($n=0$) to cause radiation at $\Omega_{{\rm{R}},0}=0.574$, and the first excited state ($n=1$) at $\Omega_{{\rm{R}},1}=0.635$.
%
%
%
%
A close up view of the resonant radiation is shown in Fig.~\ref{fig:03}(b).
%
%
The resonance spectra for trapped-states of order $n=0,1$ fit their predicted loci reasonably well.
%
%
%
The resonances for $n=2,3,4$, however, exhibit overlapping spectra.
%
%
%
%
This can be explained by radiative broadening: higher states decay faster, leading to broader resonance-linewidths.
An analysis resolving the overlapping resonance-peaks in Fig.~\ref{fig:03}(b) is provided in Appendix~\ref{appendix:d}.
%
%
%
We note that the resonance loci are restricted to frequencies satisfying $D_T(\Omega_{{\rm{R}},n})>0$, explaining the sudden cutoff of the spectrum at $\Omega_c\approx0.760$, defined by $D_{{\rm{T}}}(\Omega_c)=0$.
%
%
%
%
%
%
Increasing the duration of the soliton to $t_0=11$, i.e.\ an isotope of the previously considered meta-atom, we observe a shift of the discernible spectral lines to higher frequencies [Fig.~\ref{fig:03}(c)].
%
%
From our numerical experiments, we find that the $n=0$ resonance shifts by the amount $\Delta \Omega_{{\rm{R}},0}\approx0.026$ ($\Delta \Omega_{{\rm{R}},1}\approx0.019$).
This trend continues upon further increase of $t_0$, see Figs.~\ref{fig:03}(d,e).
In either case, the cutoff frequency remains at $\Omega_c=0.760$.
As evident from Fig.~\ref{fig:03}(d), for the $(4.525,12)$-nuclide, even four distinct spectral peaks can be distinguished.
%
%
%
%
%
The origin of the observed shift lies in the level spectra of two meta-atom isotopes, which are shifted with respect to each other, despite the fact that they have an equal number of trapped states. 
This represents an analog of the isotopic shift in atomic spectroscopy~\cite{King:BOOK:1984}.
A similar effect can be found by changing the value of $\nu$ while maintaining $N$, see Appendix~\ref{appendix:c}.
%

%
%
%
We can estimate the observed frequency shift on basis of a linear approximation of the dispersion in the vicinity of the resonances at $\Omega_0=0.58$, given by $D^\prime(\Omega) = \beta_0^\prime+ \beta_1^\prime (\Omega-\Omega_0)$ with $\beta_0^\prime=0.046$, and $\beta_1^\prime=-0.327$.
%
%
%
From the approximate phase-matching relation $-\kappa_n\approx D^\prime(\Omega_{{\rm{R}},n})$ we find $\Omega_{{\rm{R}},n} \approx \Omega_0 - v_{{\rm{R}}}(\beta_0^\prime + \kappa_{n})$.
%
%
%
Bearing in mind that $\kappa_n= \kappa_n(\nu,t_0)$, and assuming small parameter variations $\Delta t_0$ and $\Delta \nu$, yields the overall frequency shift 
$\Delta\Omega_{{\rm{R}},n} = \partial_{t_0}\Omega_{{\rm{R}},n}\, \Delta t_0 + \partial_\nu\Omega_{{\rm{R}},n}\, \Delta \nu$ resulting in the estimate 
%
%
\begin{align}
\Delta \Omega_{{\rm{R}},n}\approx 2 v_{\rm{R}} \kappa_n \left[ \Delta t_0/t_0 -\Delta \nu/(\nu-n)  \right].\label{eq:spec_shift}
\end{align}
For the $(4.525,10)$-nuclide and $\Delta t_0=1$, $\Delta \nu =0$, referring to Figs.~\ref{fig:03}(b,c), Eq.~(\ref{eq:spec_shift}) predicts the $n=0$ shift $\Delta \Omega_{{\rm{R}},0}\approx0.029$ ($\Delta \Omega_{{\rm{R}},1}\approx0.018$), providing excellent results despite the many approximations involved.
Adopting nuclear physics terminology \cite{Gold:BOOK:1997}, atoms with equal atomic number and mass number, but different physical properties, are referred to as ``isomeres''.
This technical expression applies to meta-atom nuclides with equal $t_0$, differing in their value of $\nu$ without changing $N$.
%
%
Consequently, the two terms in Eq.~(\ref{eq:spec_shift}) represent  \mbox{isotopic shift ($\propto \Delta t_0$)}, and \mbox{isomeric shift ($\propto \Delta \nu$)}.

\begin{figure}[t!]
\centering{\includegraphics[width=\linewidth]{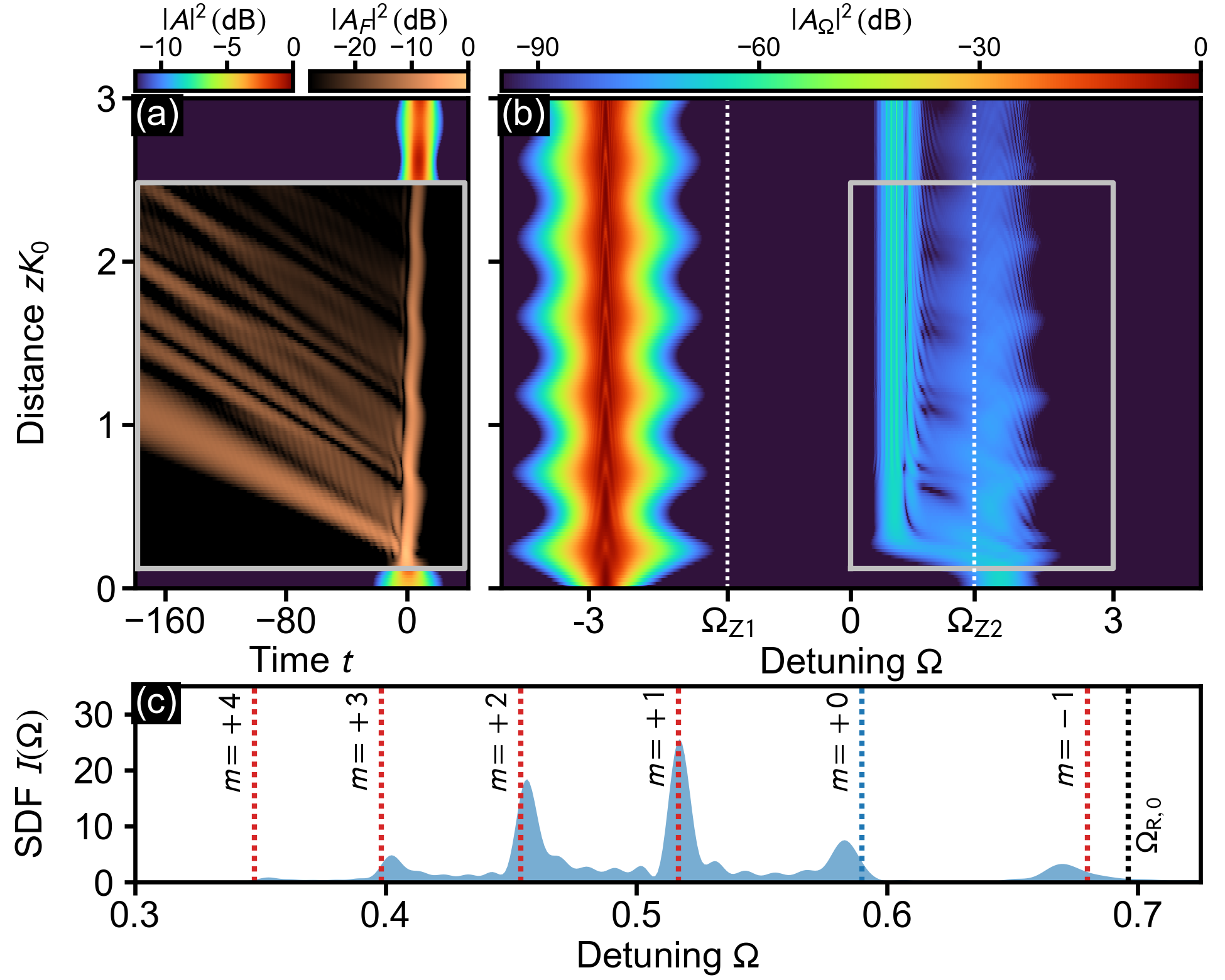}}
\caption{Splitting of spectral lines for a vibrating meta-atom.
(a) Time-domain propagation dynamics, and,
(b) spectrum, for soliton order $N_{\rm{S}}=1.5$ ($K_{\rm{S}}\approx0.026$).
Dotted lines indicate zero-dispersion points.
The field $A_F$, enclosed by the box in (a), is defined by the spectrum enclosed by the box in (b).
(c) Shifting and splitting of the groundstate resonance, in excellent agreement with the phase matching results $\Omega_{{\rm{R}},0}^m$ for $m\in\{0,\pm1, 2,3,4\}$.
$\Omega_{{\rm{R}},0}$ indicates the resonance for $N_{\rm{S}}=1$.
%
%
\label{fig:04}}
\end{figure}

%
%
In atomic physics, various mechanisms are known that yield a broadening and splitting of spectral lines \cite{King:BOOK:1984}.
%
%
Transferring a previously studied radiation mechanism for oscillating solitary waves \cite{Driben:OE:2013,Melchert:SR:2020,Melchert:NJP:2023} to the present case, we expect meta-atoms to exhibit a generic splitting mechanism for resonance-lines:
$z$-periodic amplitude and width variations of a meta-atoms soliton component give rise to a vibrating trapping potential \mbox{$V_{\rm{T}}(z,t)$~ [Eq.~(\ref{eq:VS})]};
%
%
its trapped states are subject to periodic perturbation, driving the resonant emission of multi-frequency radiation.
The result is a splitting of the resonances lines according to a modified wavenumber matching condition  $D_{\rm{T}}(\Omega_{{\rm{R}},n}^m)=-\kappa_n + m K_{\rm{S}}$, where in addition to the trapped-state index $n\in\{0,\ldots,N-1\}$ a further quantized index $m\in \{0, \pm1, \pm2, \ldots\}$ is in effect.
$K_{\rm{S}}$ is the angular wavenumber in a spatial Fourier decomposition of the soliton oscillation, and $m$ labels its $z$-harmonics. 
The range of values of $m$ is determined by the anharmonicity of the oscillation \cite{Melchert:NJP:2023}. 
%
%
%
Figure~\ref{fig:04} demonstrates this effect for the groundstate of a vibrating meta-atom.
As evident from Fig.~\ref{fig:04}(c), increasing the soliton order indeed causes a shifting as well as a splitting of spectral lines.
We note a remarkable similarity to the quantum mechanical Zeeman-effect, describing the splitting of spectral lines in presence of a static magnetic field wherein electrons acquire energy-shifts $\propto m_j \omega_L$, proportional to the Larmor frequency $\omega_L$ and magnetic quantum number $m_j$.
In our case, the splitting of resonance lines is due to wavenumber-shifts $\propto m K_{\rm{S}}$, proportional to the angular wavenumber $K_{\rm{S}}$ and spatial harmonic index $m$.
%

Finally, we point out that the portrayed association to 1D quantum physics renders the HONSE~(\ref{eq:HONSE}) also interesting from the point of view of 1D scattering problems.
In this regard we note that for integer $\nu$, the potential~(\ref{eq:VS}) is reflectionless \cite{Lekner:AJP:2007}.
%
%


\paragraph*{Conclusions.}
%
%
In conclusion, we have established an analogy between nonlinear photonic meta-atoms and 1D soft-core atoms studied in quantum dynamics.
While the bound-state structure of atoms depends on the nuclear parameters, the trapped state structure of meta-atoms depends on the solitons parameters.
%
%
Under the propagation dynamics governed by Eq.~(\ref{eq:HONSE}), higher-order dispersive effects couple the trapped-states to unique resonances, comprising the spectral fingerprint of a meta-atom.
We have demonstrated that subtle changes in the trapped-state eigenspectrum result in frequency shifts on the resonance lines.
%
%
A simplified modeling approach revealed two distinct contributions, allowing to draw an analogy to isotopic and isomeric shifts, i.e.\ spectroscopic tools to determine volume and charge distribution effects on atomic spectra, respectively.
%
%
%
Finally, we have shown a cross-phase-modulation induced mechanism that results in shifting and splitting of resonance lines for vibrating meta-atoms, inspiring an analogy to the Zeeman effect in quantum mechanics.  
%
%
We expect further broadening and splitting mechanisms to appear in the collision between meta-atoms and dispersive waves.
%
%
The reported results provide the groundwork for identifying and exploiting such phenomena in future photonic applications.

%
%
%
%
%


%
%

\vskip 0.25cm
\paragraph*{Funding --}
Deutsche Forschungsgemeinschaft (EXC 2122, projectID 390833453).
\vskip 0.25cm
\paragraph*{Data availability --} 
Data underlying the results presented in this study is openly available at \url{https://doi.org/10.5281/zenodo.18771909}.

\appendix

\begin{widetext}

\begin{figure}[t!]
\centering{\includegraphics[width=0.75\linewidth]{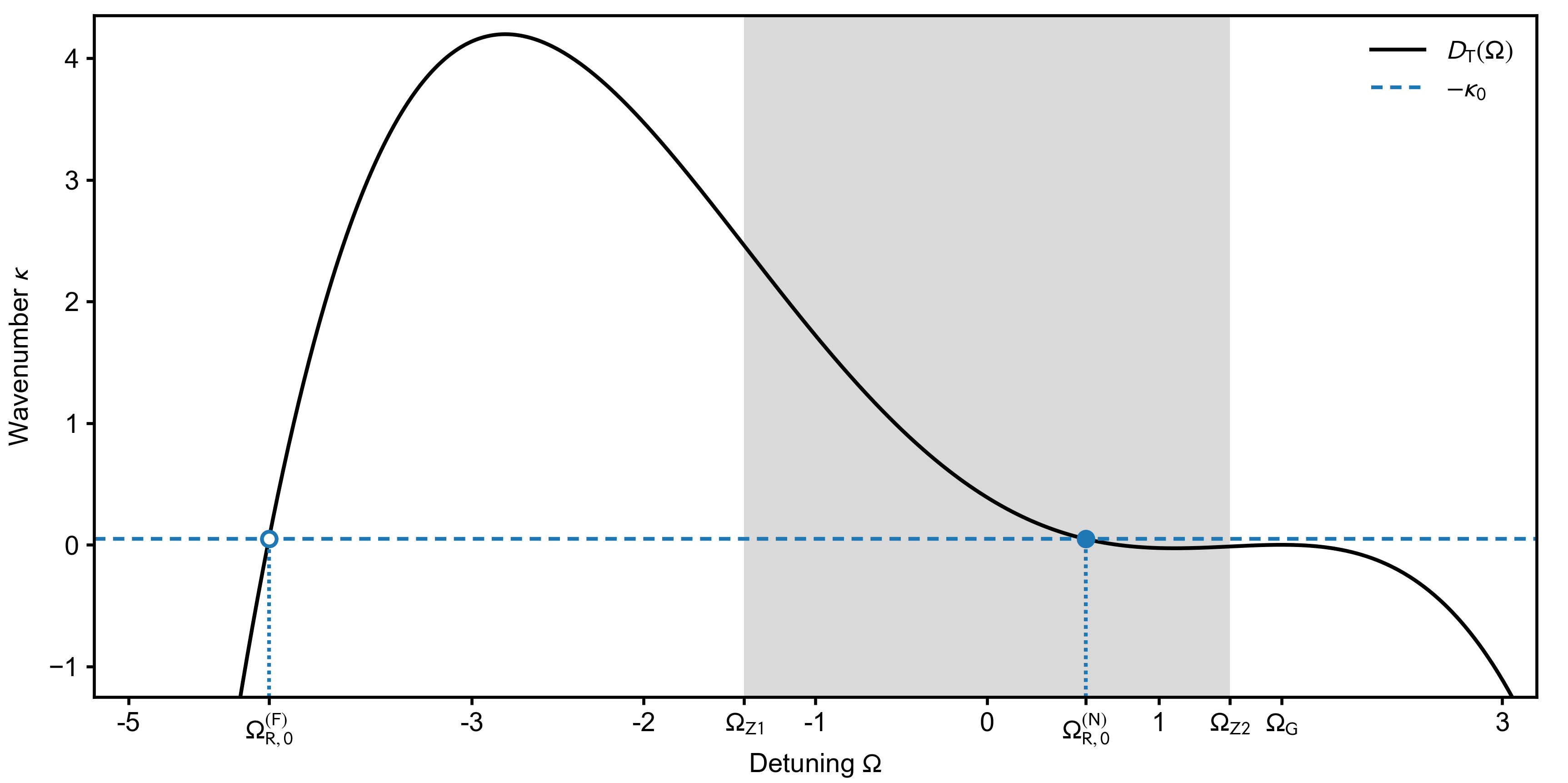}}
\caption{Details on the phase-matching resonances in Figs.~\ref{fig:03} and \ref{fig:04}.
(a) Graphical solution of the phase-matching condition~(\ref{eq:PM_condition}) for the $n\!=\!0$-resonances for the example in Fig.~\ref{fig:03} (see main document).
The horizontal dashed line indicates the sign-inverted wavenumber eigenvalue $\kappa_0\approx -0.0481$.
Filled and open circles indicate the corresponding near and far resonance loci $\Omega_{{\rm{R}},0}^{({\rm{N}})}\approx 0.574$, and  $\Omega_{{\rm{R}},0}^{({\rm{F}})}\approx -4.182$.
\label{fig:SFIG00}}
\end{figure}

\section{Details on the phase-matching analysis}
\label{appendix:a}
As discussed in the main document, the reduction of the HONSE~(\ref{eq:HONSE}), in which photonic meta-atoms emerge, is based on a separable two-pulse ansatz [Eq.~(\ref{eq:ansatz}) of the main document].
It proceeds by splitting the HONSE into a coupled system of equations for the two pulses, linearization with respect to the weak pulse $A_{\rm{G}}$ [Eqs.~(\ref{eq:coupled_set}) of the main document], and by neglecting higher-orders of dispersion local to $\Omega_{\rm{S/G}}$.
Omitting the latter simplifications results in 
%
\begin{align}
i\partial_z A_{\rm{G}} &= \left[-D_{\rm{T}}(i\partial_t)  + V_{\rm{T}}(t) -\gamma |A_{\rm{G}}|^2 \right] A_{\rm{G}}, \label{seq:GPE}
\end{align}
i.e.\ a Gross-Pitaevskii-type equation with dispersion operator
\begin{align}
D_{\rm{T}}(i\partial_t) = \frac{\beta_{2,\rm{G}}}{2}(i\partial_t)^2 + \frac{\beta_{3,\rm{G}}}{6}(i\partial_t)^3 +\frac{\beta_{4,{\rm{G}}}}{24}(i\partial_t)^4, \label{eq:DT}
\end{align}
where $\beta_{n,{\rm{G}}}=\partial_\Omega^nD|_{\Omega=\Omega_{\rm{G}}}$.
%
%
Below, we derive the wavenumber-matching relation $-\kappa_n = D_{\rm{T}}(\Omega_{{\rm{R}},n})$,
used to predict the loci $\Omega_{{\rm{R}},n}$ at which a meta-atoms trapped states couple to the continuum.
The basis of this derivation is a phase-matching analysis \cite{Yulin:OL:2004,Skryabin:PRE:2005} applied to Eq.~(\ref{seq:GPE}).

In the interest of simplicity, we consider only a single trapped state $\phi_n$ with wavenumber eigenvalue $\kappa_n$, satisfying Eq.~(\ref{eq:SE}) of the main document. 
%
%
Accounting for higher-orders of dispersion in Eq.~(\ref{eq:DT}), 
a perturbed ansatz for the solution of Eq.~(\ref{seq:GPE}) can be written as 
\begin{align}
A_{\rm{G}}(z,t) = \phi_n(t)\,e^{-i\kappa_n z} + g(z,t), \label{eq:PM_ansatz}
\end{align}
wherein  $g\equiv g(z,t)$, with $\max(|g(\cdot,t)|)\ll \max(|\phi_n(t)|)$, describes a weak radiation field emitted by the trapped state $\phi_n$.
Using Eq.~(\ref{eq:PM_ansatz}) in Eq.~(\ref{seq:GPE}), and assuming stationarity of the trapped state, we obtain a governing equation for the field $g$ in the form
\begin{align}
\left[ i\frac{\beta_{3,\rm{G}}}{6}\partial_t^3 +\frac{\beta_{4,{\rm{G}}}}{24}\partial_t^4 \right]\phi_n e^{-i\kappa_n z} = \left[ i\partial_z g + D_{\rm{T}}(i\partial_t) - V_{\rm{T}}(t)  + 2\gamma \phi_n^2 \right]g - \gamma g^*\phi_n^2 e^{-2i\kappa_n z},\label{eq:g_driving}
\end{align}
wherein the expression on the left-hand-side drives the generation of  radiation modes through the effect of higher-orders of dispersion.
%
%
To find continuum modes of Eq.~(\ref{eq:g_driving}) away from the meta-atom,  we neglect the ``localization''-terms $\propto V_{\rm{T}}$, and $\propto\phi_n^2$, and make a plane-wave ansatz $g(z,t) \propto \exp\{-i [\Omega t - D_{\rm{T}}(\Omega)z]\}$.
%
Phase-matching with the driving term on the left-hand-side is then achieved under the condition
\begin{align}
-\kappa_n \stackrel{!}{=} D_{{\rm{T}}}(\Omega).\label{eq:PM_condition}
\end{align}
Frequency detunings at which Eq.~(\ref{eq:PM_condition}) is satisfied, specify the resonance loci at which weak radiation modes are generated.

%
%
Considering the example discussed in Figs.~\ref{fig:03}(a,b) of the main document, the wavenumber-matching condition Eq.~(\ref{eq:PM_condition}) yields two $n\!=\!0$-resonances at $\Omega_{{\rm{R}},0}^{({\rm{N}})}\approx 0.574$, and  $\Omega_{{\rm{R}},0}^{({\rm{F}})}\approx -4.182$, see Fig.~\ref{fig:SFIG00}. The superindices N, for ``near'', and F, for ``far'', express the degree of proximity between the resonance location and that of the trapped state at $\Omega_{\rm{G}}$. 
%
%
In each case, as per Eq.~(\ref{eq:g_driving}), the strength of driving is proportional to the amplitude of the frequency domain representation of $\phi_0$ at $\Omega_{{\rm{R}},0}^{({\rm{N/F}})}$.
%
%
In case of $\Omega_{{\rm{R}},0}^{({\rm{N}})}$, the driving operates near to the localized trapped state, resulting in an efficient transfer of energy to resonant modes.
In case of $\Omega_{{\rm{R}},0}^{({\rm{F}})}$, the driving operates far from the localized trapped state, resulting in  negligible transfer of energy to resonant modes.
%
%
In the main document we therefore consider only the resonance nearest to $\Omega_{{\rm{G}}}$, i.e.\ $\Omega_{{\rm{R}},0}\equiv \Omega_{{\rm{R}},0}^{({\rm{N}})}$, and neglect $\Omega_{{\rm{R}},0}^{({\rm{F}})}$.

%
When the trapping-potential $V_{\rm{T}}$ exhibits (slow) $z$-periodic variations of depth and width, as, e.g., induced by a non-fundamental soliton considered in Fig.~\ref{fig:04} of the main manuscript, $\phi_n$ and $\kappa_n$ will exhibit a corresponding (slow) periodicity. 
In such a situation, as discussed in the context of two-color soliton molecules \cite{Oreshnikov:PRA:2022,Melchert:NJP:2023}, and models with $z$-periodic group-velocity dispersion \cite{Conforti:SR:2015}, 
the effective driving-wavenumber in Eq.~(\ref{eq:g_driving}) will reflect this $z$-periodicity.
Consequently, the modified phase-matching condition 
\begin{align}
-\kappa_n + m K_{\rm{S}} \stackrel{!}{=} D_{\rm{T}}(\Omega), \quad  \text{for}\quad m\in \{0, \pm1, \pm2, \ldots\}, \label{eq:PM_condition_vib}
\end{align}
can be derived, wherein $K_{\rm{S}}$ is the angular wavenumber in a spatial Fourier decomposition of the $z$-periodic oscillation, and $m$ labels the corresponding $z$-harmonics \cite{Oreshnikov:PRA:2022}.

\section{Details on the split-step Fourier method}
\label{appendix:b}

%
%
For our numerical experiments we use an implementation of the fixed-stepsize ``symmetric split-step Fourier method'' (SSFM) \cite{DeVries:APL:1987,Melchert:CPC:2022,Melchert:OL:2021}, as well as the adaptive-stepsize ``conservation quantity error method'' (CQE) with step-size selection governed by photon number conservation \cite{Heidt:JLT:2009,Melchert:CPC:2022}.
The FFSM was used in the numerical experiments reported in Figs.~\ref{fig:02} and \ref{fig:03} of the main manuscript, while the CQE was used in those reported in Figs.~\ref{fig:04} and \ref{fig:SFIG02}.

%
%
The fixed stepsize $h$ of our SSFM is chosen according to a pulse-walk-off argument \cite{Sinkin:JLT:2003}.
The fastest temporal walk-off in the considered propagation scenario is determined by the soliton at detuning $\Omega_{\rm{S}}$ and the resonant radiation, here represented by the detuning $\Omega_{{\rm{R}},0}$.
The corresponding GV mismatch reads $\Delta \beta_1 = |D_1(\Omega_{\rm{S}})-D_1(\Omega_{{\rm{R}},0})|$.
The stepsize is then taken as $h=C\,t_0/ \Delta \beta_1$, with soliton duration $t_0$, and $C=10^{-3}$. 
The dimensionless parameter $C$ specifies the allowed walk-off per propagation step in units of the soliton duration. 
We ensured that results did not change upon further decrease of $h$.
%
%
For the propagation scenario in Fig.~\ref{fig:04} of the main manuscript, the resulting value reads $h\approx 0.046$ (for $t_0=15$ and $\Delta \beta_1 \approx 0.33$).
In comparison, for the same propagation scenario, the CQE with local goal-error $\delta_G=10^{-12}$ achieves a minimum stepsize $\min[h(z)]\approx 0.040$, see Fig.~\ref{fig:SFIG01}.

\begin{figure}[t!]
\centering{\includegraphics[width=0.75\linewidth]{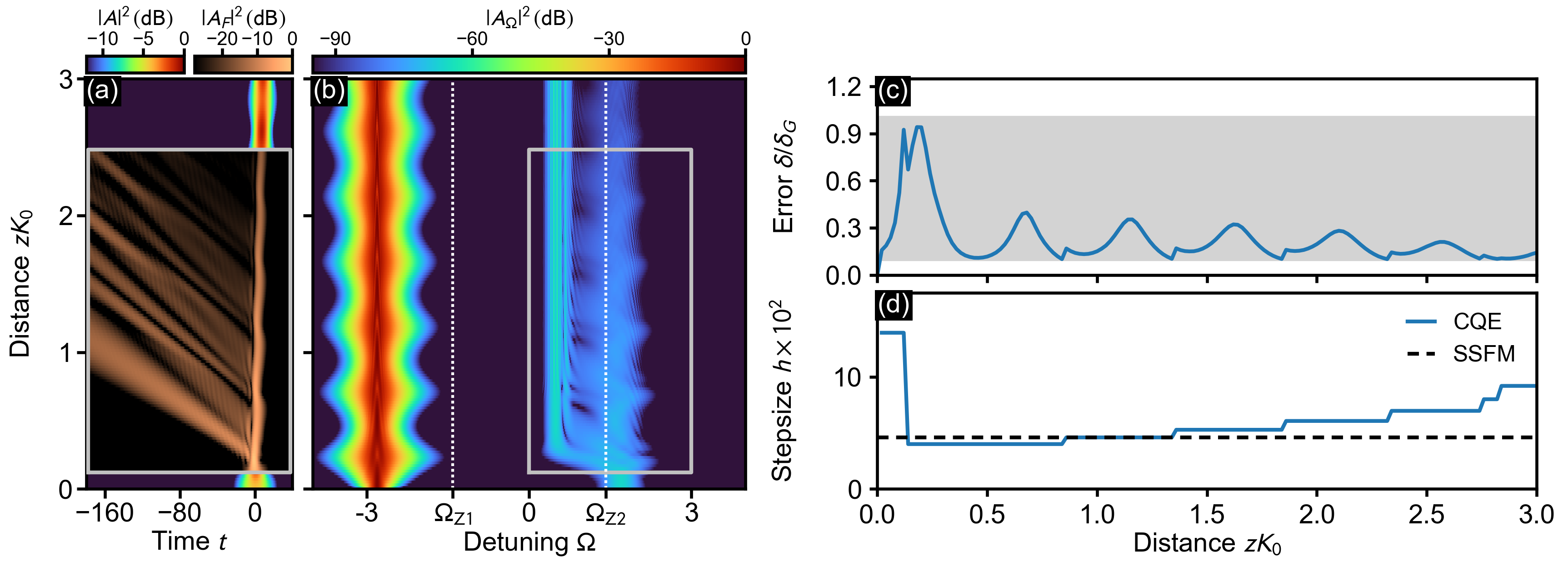}}
\caption{Performance of the CQE for the propagation of a vibrating meta-atom.
(a) Time-domain propagation dynamics, and, (b) corresponding spectrum.
(b) Variation of the relative photon number error.
Shaded area highlights the goal error range $[0.1\delta_G, \delta_G]$.
(c) Variation of the local stepsize $h(z)$ (labeled CQE), and fixed stepsize of the SSFM method (labeled SSFM).
\label{fig:SFIG01}}
\end{figure}

\section{Isomeric shift on the spectral signature}
\label{appendix:c}

We demonstrate the isomeric shift on the resonance spectra of meta-atom isotopes in Fig.~\ref{fig:SFIG02}. 
We consider the propagation scenario discussed in the main text with soliton duration $t_0=10$, and soliton center frequencies $-2.794<\Omega_{\rm{S}}<-2.812$, selected in a range in which the number of trapped states is equal to $N=4$.
All spectra are extracted at propagation distance $z=10\,K_0^{-1}$, and are normalized as $\int_{0.5}^{0.9}I(\Omega)~{\rm{d}}\Omega=1$.
%
In this case, the choice of $\Omega_{\rm{S}}$ affects the location of $\Omega_{\rm{G}}$ and thus the local dispersion profile $D_{\rm{T}}(\Omega)$.
Therefore, the cutoff frequency $\Omega_c$, defined through $D_{\rm{T}}(\Omega_c)=0$, also depends on $\Omega_{\rm{S}}$, see Fig.~\ref{fig:SFIG02}.

\begin{figure}[t!]
\centering{\includegraphics[width=0.5\linewidth]{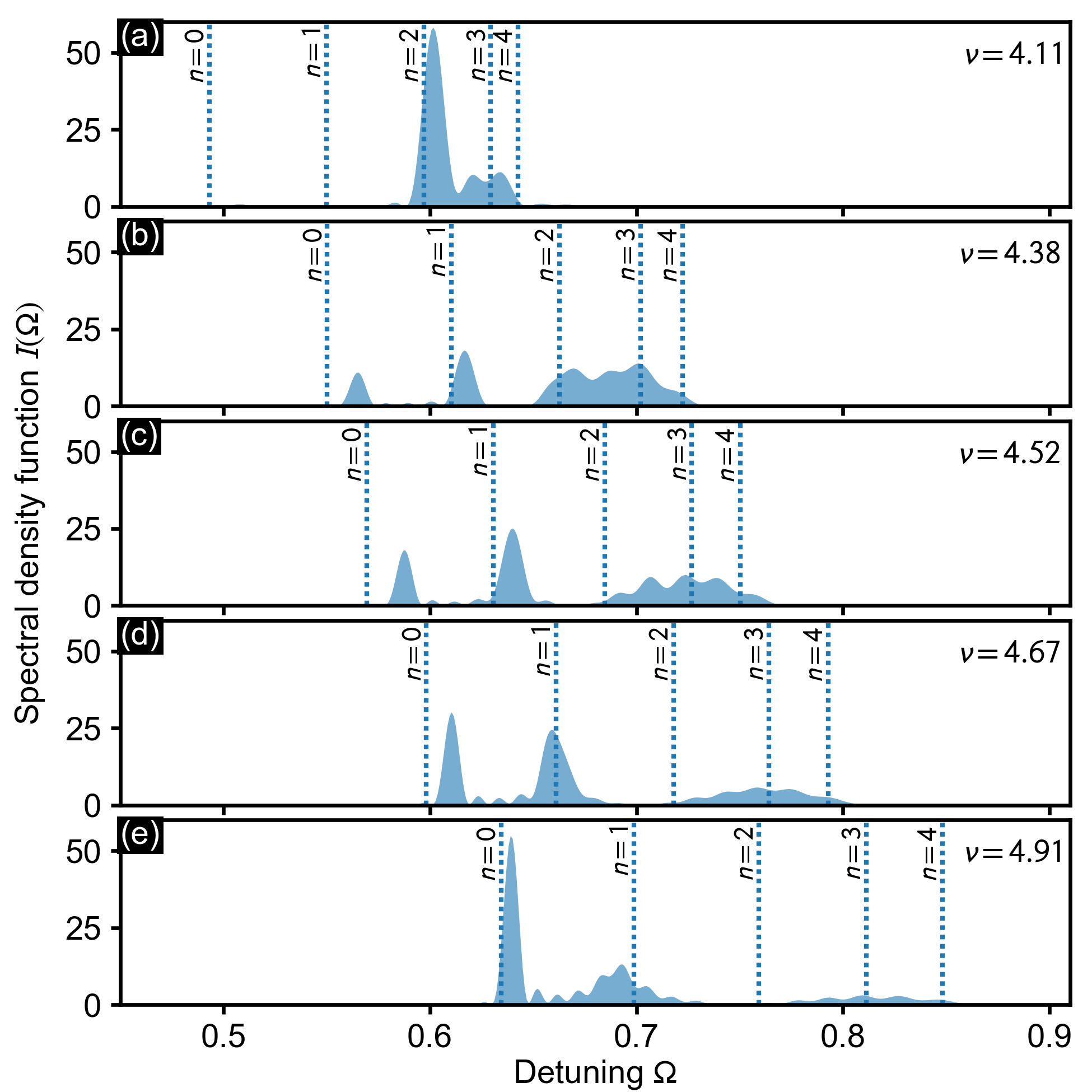}}
\caption{Isomeric shift of resonance lines for meta-atom isotopes.
Spectral density function for fixed $t_0=10$, and selected values of the meta-atom center frequency $\Omega_{\rm{S}}$, giving rise to an effective charge $4<\nu<5$.
(a) $\Omega_{\rm{S}}=-2.797$ with $\nu=4.11$, 
(b) $\Omega_{\rm{S}}=-2.803$ with $\nu=4.38$, 
(c) $\Omega_{\rm{S}}=-2.805$ with $\nu=4.52$, 
(d) $\Omega_{\rm{S}}=-2.808$ with $\nu=4.67$, 
(e) $\Omega_{\rm{S}}=-2.812$ with $\nu=4.91$, 
Dashed lines indicate the predicted resonances at $\Omega_{{\rm{R}},n}$ for $n=0,\ldots,4$.
%
%
\label{fig:SFIG02}}
\end{figure}

\section{Peak-fitting analysis of the resonance spectrum in Fig.~3(b)}
\label{appendix:d}

In Fig.~\ref{fig:03}(b) of the main manuscript, the $(4.525,10)$--nuclids resonance spectrum was shown to exhibit overlapping spectral peaks associated to the trapped states of order $n=2$ through $4$.
The resulting spectral density function (SDF) $I(\Omega)$ can be analyzed by standard peak-fitting techniques \cite{Major:JVST:2020}, providing means to extract further quantitative information from the meta-atoms \mbox{(numerical-)spectroscopy} data.

Below we report such a peak-fitting analysis, assuming Gaussian lineshapes 
\begin{align}
G(\Omega;I_0,\Omega_c,\sigma) = I_0\,\exp[-(\Omega-\Omega_c)^2/2\sigma^2], \label{eq:lineshape}
\end{align}
with parameters specifying the peak height ($I_0$), loccation ($\Omega_c$), and, width ($\sigma$).
Since the component peaks for $n=0$ and $n=1$ are well separated, we perform an independent fit for each component within a narrow detuning-interval surrounding its peak.
%
For the peaks corresponding to $n=2,3,4$, we instead perform a combined fit within a larger detuning-interval, indeed allowing to resolve the individual component-peaks within the overlapping spectrum. 
As evident from Fig.~\ref{fig:SFIG04}, the total fit-function $G_{tot}(\Omega)$, representing a sum of all five component-peaks, excellently fits the SDF $I(\Omega)$.
%
%
%
%
%
We further point out that the increasing peak-width observed for increasing component-number $n$ meets the expectation that trapped-states of higher order decay faster, implying increasing resonance widths \cite{Zhang:EL:2026}. 

\begin{figure}[t!]
\centering{\includegraphics[width=0.75\linewidth]{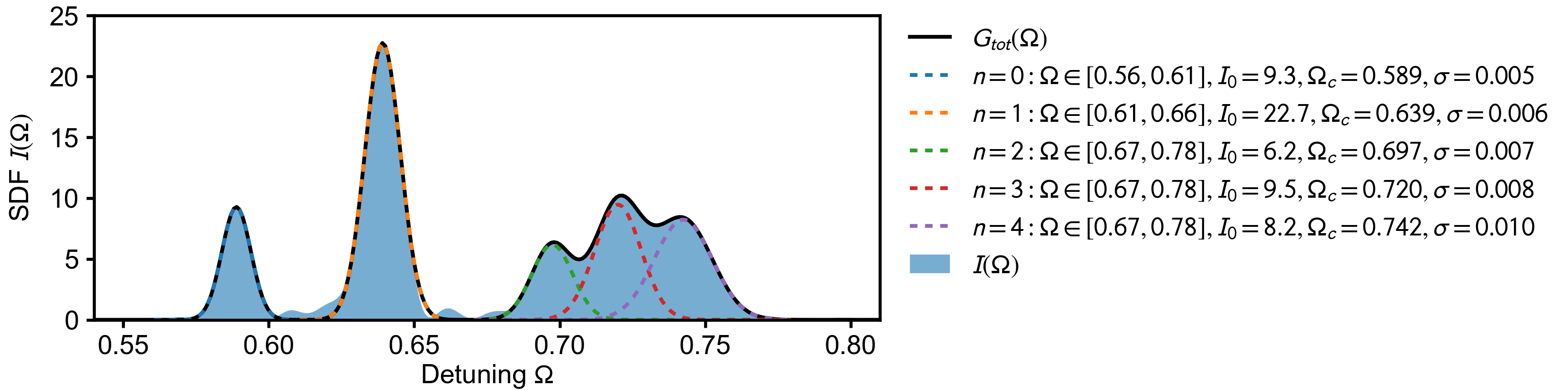}}
\caption{Peak-fitting analysis of the SDF of Fig.~\ref{fig:03}(b).
For each component peak ($n=0,\ldots,4$), the key lists the considered fit-range ($[\Omega_{\rm{min}},\Omega_{\rm{max}}]$), peak intensity ($I_0$), location ($\Omega_c$), and width ($\sigma$).
%
\label{fig:SFIG04}}
\end{figure}

\section{Code ocean compute capsule}
\label{appendix:e}

To facilitate a reproduction of the numerical experiments reported in the main document, we provide a Code Ocean \cite{Clyburne:CO:2019} compute capsule \cite{IsotopicShift:CO:2026}.
It implements the propagation dynamics resulting from the initial condition 
\begin{align}
A_0(t)=\sqrt{P_0}\,{\rm{sech}}(t/t_0)\,e^{-i\Omega_{\rm{S}}t} + 10^{-3}\sqrt{P_{0}} \exp(-t^2/2t_{{\rm{G}}}^2)\,e^{-i\Omega^\prime t},
\end{align}
consisting of a soliton with parameters $t_0=10$, $P_0\approx 0.03$, at  $\Omega_{\rm{S}}=-2.81$, and a superimposed weak Gaussian pulse with $t_{{\rm{G}}}=10$ at $\Omega^\prime=1.72$, governed by the HONSE~(1).
The provided code reproduces Fig.~\ref{fig:SFIG03}, showing the propagation scenario of Fig.~\ref{fig:02} (see main document), and demonstrating the phase-matching analysis of Fig.~\ref{fig:03} (see main document).

\begin{figure}[t!]
\centering{\includegraphics[width=0.75\linewidth]{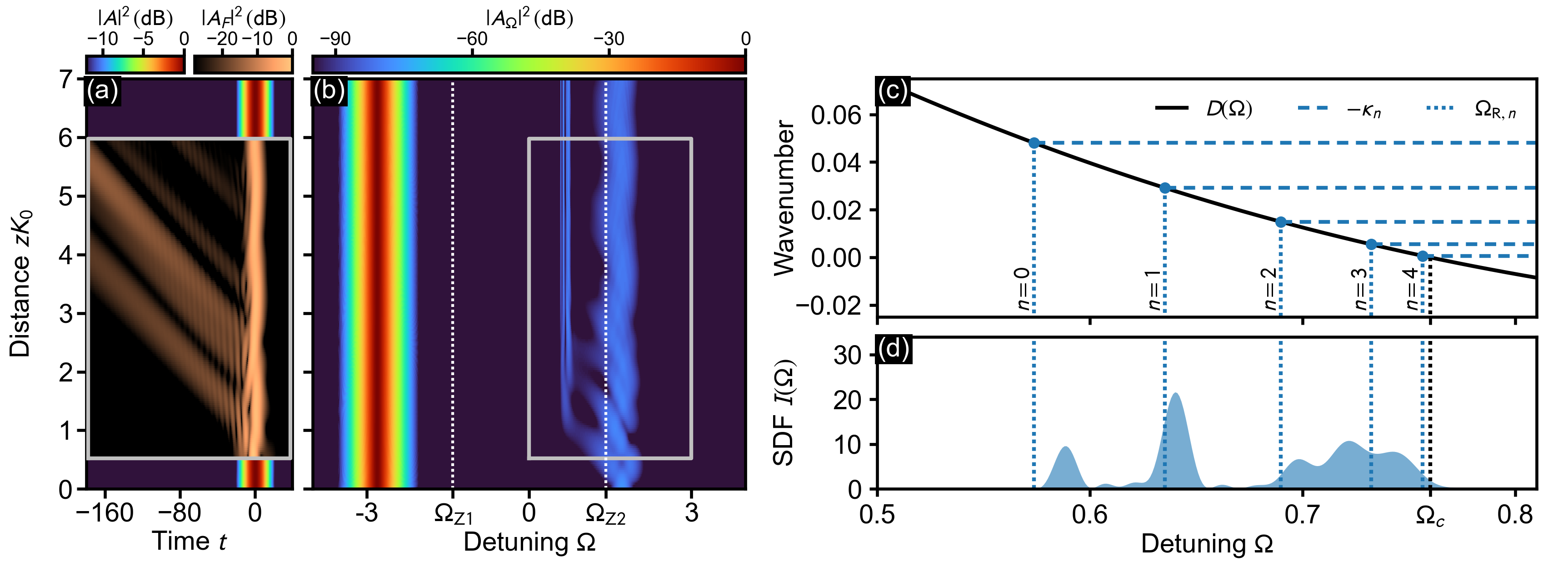}}
\caption{Figure generated by the code-ocean compute capsule.
(a) Time-domain propagation dynamics of a soliton, and a superimposed Gaussian pulse (cf.~Fig.~\ref{fig:02} of the main document),
with propagation distance scaled by the wavenumber $K_0$.
(b) Corresponding spectrum.
Dotted lines indicate zero-dispersion points.
The field $A_F$, enclosed by the box in (a), is defined by the spectrum enclosed by the box in (b).
(c) Graphical solution of the phase matching analysis, and, (d) spectral density function (SDF) emphasizing the resonance spectrum (cf.~Fig.~\ref{fig:03} of the main document).
\label{fig:SFIG03}}
\end{figure}

\end{widetext}

\bibliography{references}

\end{document}